\documentclass[review,onefignum,onetabnum]{article}
\pdfoutput=1

\usepackage{latexsym}
\usepackage{amssymb,amsbsy,amsmath,amsfonts,amssymb,amscd,amsthm}
\usepackage{subfigure}
\usepackage{graphicx}
\usepackage{hyperref}
\usepackage{dsfont}
\usepackage{mathtools}
\usepackage{pdfsync}
\usepackage{epsfig,epstopdf}
\usepackage{caption}
\usepackage{xcolor}
\usepackage{hyperref}
\usepackage{float}

\newtheorem{rmrk}[]{Remark}

\newcommand{\beq}{\begin{equation}}
\newcommand{\eeq}{\end{equation}}

\def\red#1{\textcolor{black}{#1}}

\newcommand {\e}  {\varepsilon}

\newcommand{\pd}{\partial}
\newcommand{\dt}{\pd_{t}}
\newcommand{\dr}{\nabla_{\bf r}}
\newcommand{\dx}{\pd_{x}}

\newcommand{\dxx}{\pd^{2}_{xx}}

\newcommand{\nep}{n_{\e}}

\newcommand{\ue}{u_{\e}}
\newcommand{\xb}{\bar{x}}

\newcommand{\xbi}{\xb^{\infty}}
\newcommand{\ii}{I^{\infty}}

\DeclareMathOperator*{\argmax}{arg\,max}
\usepackage{amsopn}

\title{Modelling the emergence of phenotypic heterogeneity in vascularised tumours}

\author{Chiara Villa\thanks{University of St Andrews, UK 
  (cv23@st-andrews.ac.uk)}
\and
Mark A. J. Chaplain\thanks{University of St Andrews, UK 
  (majc@st-andrews.ac.uk)}
\and
Tommaso Lorenzi\thanks{Politecnico di Torino, IT 
  (tommaso.lorenzi@polito.it)}
}

\begin{document}
\date{}
\maketitle

\begin{abstract}
We present a mathematical study of the emergence of phenotypic heterogeneity in vascularised tumours. Our study is based on formal asymptotic analysis and numerical simulations of a system of non-local parabolic equations that describes the phenotypic evolution of tumour cells and their nonlinear dynamic interactions with the oxygen, which is released from the intratumoural vascular network. Numerical simulations are carried out both in the case of arbitrary distributions of intratumour blood vessels and in the case where the intratumoural vascular network is reconstructed from clinical images obtained using dynamic optical coherence tomography. The results obtained support a more in-depth theoretical understanding of the eco-evolutionary process which underpins the emergence of phenotypic heterogeneity in vascularised tumours. In particular, our results offer a theoretical basis for empirical evidence indicating that the phenotypic properties of cancer cells in vascularised tumours vary with the distance from the blood vessels, and establish a relation between the degree of tumour tissue vascularisation and the level of intratumour phenotypic heterogeneity.\end{abstract}

\section{Introduction}
Spatial variability in the intratumoural concentration of oxygen plays a pivotal role in the emergence and development of phenotypic heterogeneity among tumour cells~\cite{alfarouk2013riparian,gillies2012evolutionary,marusyk2012intra,molavian2009fingerprint,sun2015intra}. This is exemplified in a growing body of experimental and clinical studies demonstrating that tumour cells with different phenotypic properties occupy tumour regions which are characterised by different oxygen levels. In particular, hypoxic parts of the tumour (\emph{i.e.} regions where oxygen levels are below normal physiological levels) are mainly populated by slow-dividing cells, which display higher levels of hypoxia-inducible factors, such as HIF-1~\cite{giatromanolaki2001relation,lee2004hypoxia,padhani2007imaging,semenza2003targeting,strese2013effects,tannock1968relation,zhao2013targeting}. 
On the other hand, fast-dividing cells with lower levels of expression of hypoxia-inducible factors are primarily detected in well-oxygenated parts of the tumour tissue (\emph{i.e.} the tumour border in avascular tumours and the regions in the vicinity of blood vessels in vascularised tumours)~\cite{giatromanolaki2001relation,tannock1968relation,semenza2003targeting}. 
This impinges on anti-cancer treatment by making \red{it} impossible for single biopsies to exhaustively portray the phenotypic composition of the whole tumour tissue~\cite{burrell2014tumour,poleszczuk2015evolution,yap2012intratumor}.

Previous empirical and theoretical work has suggested that this may be the outcome of eco-evolutionary dynamics driven by interactions between oxygen molecules and tumour cells~\cite{alfarouk2013riparian,gallaher2013evolution,gay2016tumour,loeb2001mutator,marusyk2012intra,michor2010origins}. In particular, it has been hypothesised that the nonlinear interplay between impaired oxygen delivery caused by structural abnormalities present in the tumour vasculature~\cite{fukumura2010tumor,jain1988determinants,jordan2012targeting,padhani2007imaging,vaupel1989blood}, limited oxygen diffusion and oxygen consumption by tumour cells may lead to the creation of distinct ecological niches in which tumour cells with different phenotypic characteristics can be selected~\cite{alfarouk2013riparian,casciari1992variations,gatenby2007cellular,hockel2001tumor,ibrahim2017defining,lloyd2016darwinian}. 
 
In this paper, we use a spatially explicit phenotype-structured model to elucidate the eco-evolutionary dynamics that underpin the emergence of phenotypic heterogeneity in vascularised tumours. Building upon the modelling framework that we developed in~\cite{lorenzi2018role,lorz2015modeling}, the model comprises a non-local parabolic partial differential equation (PDE) that governs the local phenotypic distribution of cells within the tumour tissue. Similar PDEs modelling the evolutionary dynamics of space- and phenotype-structured populations have recently received increasing attention from the mathematical community -- see for instance ~\cite{alfaro2017effect,alfaro2013travelling,arnold2012existence,bouin2014travelling,bouin2012invasion,bouin2015hamilton,calvez2018non,domschke2017structured,hodgkinson2018signal,jabin2016selection,mirrahimi2015asymptotic}. 

The equation for the evolution of the local phenotypic distribution of tumour cells is coupled with a parabolic PDE that governs the local concentration of oxygen, whereby a spatially heterogeneous source term captures the presence of intratumoural blood vessels which bring oxygen into the tumour tissue. Different possible definitions of such a source term are considered, including definitions that are derived from clinical images obtained using dynamic optical coherence tomography (D-OCT)~\cite{schuh2017imaging} -- \emph{i.e.} a non-invasive imaging technique that enables the visualisation of cutaneous microvasculature in 2D tissue sections with a width of and at a depth of up to several millimetres~\cite{olsen2018advances}.

Compared to previous related studies~\cite{lorenzi2018role,lorz2015modeling}, our model takes into account the effect of movement and phenotypic variation of tumour cells and, in addition, it does not rely on a quasi-stationary equilibrium assumption for the oxygen concentration. Furthermore, while these previous studies are mainly focused on avascular tumours, in this paper we consider vascularised tumours and systematically assess the impact of the degree of tumour tissue vascularisation on the level of phenotypic heterogeneity of tumour cells, which is mathematically quantified through suitable diversity indices. Moreover, numerical solutions of the model equations are here integrated with the results of formal asymptotic analysis, in order to achieve more robust and precise biological conclusions. Taken together, these elements of novelty widen considerably the range of application of the results of our study and support a more in-depth theoretical understanding of the eco-evolutionary process which leads to the emergence of phenotypic heterogeneity in vascularised tumours.

The paper is organised as follows. In Section~\ref{sec:model}, we introduce the equations of the model and the underlying modelling assumptions. In Section~\ref{sec:analysis}, we carry out a formal asymptotic analysis of evolutionary dynamics. In Section~\ref{sec:numsol}, we present a sample of numerical solutions that confirm the results of our formal analysis, and we discuss their biological implications. Section~\ref{sec:disc} concludes the paper and provides a brief overview of possible research perspectives.

\section{Description of the model}
\label{sec:model}
We model the evolutionary dynamics of cancer cells in a region of a vascularised tumour, which is approximated as a bounded set $\Omega \subset \mathbb{R}^d$ with smooth boundary $\partial \Omega$, where $d=1,2,3$ depending on the biological scenario under study. The spatial position of the tumour cells is described by a vector ${\bf r} \in \Omega$, while the  phenotypic state of the cells is modelled by a scalar variable $x \in [0,1] \subset \mathbb{R}$, which represents the normalised level of expression of a hypoxia-inducible factor. 
The phenotypic distribution of tumour cells at time $t \geq 0$ and position ${\bf r}$ is described by the function $n(t,{\bf r},x)$, while the function $s(t,{\bf r})$ describes the oxygen concentration at time $t$ and position ${\bf r}$. 

At each time $t$, we define the local cell density and the local mean phenotypic state of tumour cells, respectively, as
\begin{equation}\label{Imu}
I(t,{\bf r}) := \int_0^1  n(t,{\bf r},x) \, {\rm d}x \quad \text{and} \quad X(t,{\bf r}) := \frac{1}{I(t,{\bf r})} \int_{0}^1 x \, n(t,{\bf r},x) \, {\rm d}x.
\end{equation}
Moreover, we define the total cell number and the fraction of cells in the phenotypic state $x$ within the tumour, respectively, as
\begin{equation}\label{NF}
N(t) := \int_{\Omega} I(t,{\bf r}) \, {\rm d}{\bf r} \quad \text{and} \quad F(t,x) := \frac{1}{N(t)} \int_{\Omega} n(t,{\bf r},x) \, {\rm d}{\bf r}.
\end{equation}

\subsection{Dynamic of tumour cells}
Tumour cells divide, die, move randomly (\emph{i.e.} undergo undirected, spontaneous migration) and undergo spontaneous epimutations, that is, heritable phenotypic changes that occur randomly due to non-genetic instability and are not induced by any selective pressure~\cite{huang2013genetic}. The dynamic of the local cell phenotypic distribution $n(t,{\bf r},x)$ is governed by the following boundary value problem subject to a suitable initial condition
\beq
\label{eq:n}
\begin{cases}
\displaystyle{\partial_{t} n = \underbrace{R\big(x,I(t,{\bf r}),s(t,{\bf r})\big) n}_{\mbox{\scriptsize{cell division \& death}}} \quad + \; \underbrace{\alpha \, \partial^2_{xx} n}_{\substack{\mbox{\scriptsize{spontaneous}}\\\mbox{\scriptsize{epimutations}}}} \; + \; \underbrace{\beta \, \Delta_{\bf r} n}_{\substack{\mbox{\scriptsize{random}}\\\mbox{\scriptsize{movement}}}}} \quad \text{in } \Omega \times (0,1),
\\\\
\displaystyle{I(t,{\bf r}) :=  \int_{0}^{1} n(t,{\bf r},x) \, {\rm d}x}, 
\\\\
\partial_x n(\cdot,\cdot,0) = \partial_x n(\cdot,\cdot,1) = 0,
\\\\
\nabla_{{\bf r}} n \cdot {\bf \hat{u}} = 0 \quad \text{on } \partial\Omega,
\end{cases}
\eeq
where ${\bf \hat{u}}$ is the unit normal to $\partial\Omega$ that points outward from $\Omega$.

The first diffusion term on the right-hand side of the non-local parabolic equation~\eqref{eq:n}$_1$ describes the effect of changes in the local phenotypic distribution due to spontaneous epimutations, which occur at rate $\alpha$~\cite{chisholm2015emergence, lorenzi2016tracking}. The second diffusion term models the effect of cell random movement and the parameter $\beta$ represents the cell motility. The function $R\big(x,I(t,{\bf r}),s(t,{\bf r})\big)$ models the fitness of tumour cells in the phenotypic state $x$ at position ${\bf r}$ and time $t$ under the local environmental conditions given by the cell density $I(t,{\bf r})$ and the oxygen concentration $s(t,{\bf r})$ (\emph{i.e.} the phenotypic fitness landscape of the tumour). In particular, we define the function $R$ as
\begin{equation}\label{def:R}
R\big(x,I,s\big) := p\big(x,s\big) - \kappa \, I.
\end{equation}
Definition~\eqref{def:R} models a biological scenario whereby tumour cells die at position ${\bf r}$ and time $t$ due to competition for limited space at rate $\kappa \, I(t,{\bf r})$, with the parameter $\kappa$ being related to the local carrying capacity of the tumour. Moreover, the function $p\big(x,s(t,{\bf r})\big)$ is the division rate of cells in the phenotypic state $x$ at position ${\bf r}$ and time $t$ under the oxygen concentration $s(t,{\bf r})$. Building upon the modelling strategy presented in~\cite{lorenzi2018role}, we define the cell division rate as 
\begin{equation}\label{def:p}
p(x,s) :=  f(x) +  g(x,s).
\end{equation}
In~\eqref{def:p}, the function $g(x,s(t,{\bf r}))$ represents the rate of cell division via aerobic pathways under the oxygen concentration $s(t,{\bf r})$, while the function $f(x)$ models the rate of cell division via anaerobic pathways. Based on the biological evidence and ideas presented in~\cite{alfarouk2013riparian, ardavseva2019evolutionary, gordan2007hif, lloyd2016darwinian, xia2014knockdown}, we let cells with a lower level of expression of the hypoxia-inducible factor (\emph{i.e.} cells in phenotypic states $x \to 0$) be characterised by a higher rate of cell division via aerobic pathways, while we assume cells with a higher level of expression of the hypoxia-inducible factor (\emph{i.e.} cells in phenotypic states $x \to 1$) to be characterised by a higher rate of cell division via anaerobic pathways. Therefore, we assume the functions $f(x)$ and $g(x,s)$ to be smooth and such that
\beq
f(0) = 0, \quad f'(x) > 0 \quad \forall \, x \in [0,1),
\label{assf}
\eeq
\beq
g(1,s) = 0 \quad \forall \, s \in [0,\infty), \quad \partial_x g(x,s) < 0 \quad \forall \, (x,s) \in (0,1] \times (0,\infty).
\label{assg1}
\eeq
Moreover, we make the following natural assumptions
\beq
g(x,0) = 0 \quad \forall x \in [0,1], \quad \partial_s g(x,s) > 0 \quad \forall \, (x,s) \in [0,1) \times (0,\infty).
\label{assg2}
\eeq


\subsection{Dynamic of oxygen}
We let oxygen enter the tumour through intratumoural blood vessels, diffuse in space, decay over time and be consumed by tumour cells which divide via aerobic pathways. In this scenario, the dynamic of the oxygen concentration $s(t,{\bf r})$ is governed by the following boundary value problem 
\beq
\begin{cases}
\partial_{t}s = \underbrace{\beta_s \Delta_{\bf r} s}_{\mbox{\scriptsize{diffusion}}} \, - \, \eta_s \underbrace{\int_{0}^1 \hspace{-0.1cm} g(x,s)  \, n(t,{\bf r},x) \, {\rm d}x}_{\mbox{\scriptsize{consumption by tumour cells}}} \; - \, \underbrace{\lambda_s s}_{\mbox{\scriptsize{decay}}} \;\; + \, \underbrace{q(t,{\bf r})}_{\substack{\mbox{\scriptsize{inflow from}}\\\mbox{\scriptsize{blood vessels}}}} \quad \text{in } \Omega, 
\\\\
\nabla_{\bf r} s \cdot {\bf \hat{u}} = 0 \quad \text{on } \partial\Omega,
\end{cases}
\label{eq:s}
\eeq
subject to a suitable initial condition and coupled to the non-local parabolic equation~\eqref{eq:n}$_1$. 

In the parabolic equation~\eqref{eq:s}$_1$, the parameter $\beta_s$ is the oxygen diffusion coefficient, 
$\eta_s$ is a conversion factor linked to the rate of oxygen consumption by tumour cells, 
$\lambda_s$ is the natural decay rate of oxygen, and the source term $q(t,{\bf r})$ models the influx of oxygen from the intratumoural blood vessels. We let $\omega \subset \Omega$ be the set of points within the tumour tissue which are occupied by blood vessels and, since we do not consider the formation of new blood vessels, we assume $\omega$ to be given and remain constant in time. Therefore, we use the following definition
\beq
\label{q}
q(t,{\bf r}) := S(t,{\bf r}) \, {\bf 1}_{\omega}({\bf r}) 
\eeq
where ${\bf 1}_{\omega}$ is the indicator function of the set $\omega$ and $S(t,{\bf r})$ is the rate of inflow of oxygen through intratumoural blood vessels at position ${\bf r} \in \omega$ and time $t$.
\begin{rmrk}
In this paper, we do not take into account the effect of mechanical interactions between tumour cells and blood vessels and we do not allow tumour cell to extravasate. Therefore, focussing on the case of intratumoural blood vessels of small size, we implicitly make the following simplifying assumptions: (i) a point ${\bf r}$ can be simultaneously occupied by blood vessels and tumour cells; (ii) cell movement is not affected by the presence of blood vessels. Therefore, we do not impose any condition on $n(t,{\bf r},x)$ in $\omega$.
\end{rmrk}

\section{Formal analysis of evolutionary dynamics}
\label{sec:analysis}
In order to obtain an analytical description of the evolutionary dynamics of tumour cells, 
in this section we carry out a formal asymptotic analysis of the behaviour of the solution to the problem given by~\eqref{eq:n} subject to a suitable initial condition, under the assumption
\beq
\label{assbars}
s(t,{\bf r}) \equiv s^{\infty}({\bf r}),
\eeq
where $s^{\infty}({\bf r})$ is a given smooth positive function. We note that assumptions~\eqref{assf}-\eqref{assg2} ensure that 
\beq
0 < p(x,s) < \infty \quad \forall \, (x,s) \in [0,1] \times (0,\infty).
\label{assp0}
\eeq
Moreover, focussing on an intratumour phenotypic fitness landscape $R\big(x,I,s^{\infty}\big)$ defined according to~\eqref{def:R} with a single peak at each position ${\bf r}$ and time $t$, we let the cell division rate $p$ be a smooth function that satisfies the additional concavity assumption
\beq
\partial^2_{xx} p(x,s) < 0 \quad \forall \, (x,s) \in [0,1] \times (0,\infty).
\label{assp}
\eeq

Typical values of the epimutation rate $\alpha$ are one or two orders of magnitude larger than the rate of somatic DNA mutation \cite[p.45]{doerfler2006dna}, which is about $10^{-12} s^{-1}$~\cite{duesberg2000explaining}, and typical values of the cell diffusivity $\beta$ are about $10^{-12} \, cm^{2}s^{-1}$~\cite{smith2004measurement,wang2009mathematical}. 
Hence, spontaneous epimutations and random cell movement occur on slower time scales compared to cell division and death. To capture this fact, we introduce a small parameter $\e>0$ and assume both $\alpha := \e^2$ and $\beta := \e^2$.

Following previous studies on the long-time behaviour of non-local PDEs and integro-differential equations modelling the dynamics of continuously structured populations~\cite{barles2009concentration,chisholm2016effects,desvillettes2008selection,diekmann2005dynamics,jabin2016selection,lorz2011dirac,mirrahimi2015asymptotic,perthame2008dirac}, we use the time scaling $t \mapsto \frac{t}{\e}$ in the balance equation~\eqref{eq:n}$_1$ complemented with~\eqref{assbars}, which gives the following non-local PDE for the local cell phenotypic distribution $n\left(\frac{t}{\e},{\bf r},x\right)=n_{\e}(t,{\bf r},x)$
\beq
\label{eq:ne}
\displaystyle{\e \, \dt\nep= R\big(x,I_{\e}(t,{\bf r}),s^{\infty}({\bf r})\big) \, \nep +\e^2 \, \dxx\nep + \e^2 \, \Delta_{\bf r}\nep}.
\eeq
Considering the asymptotic regime $\e \rightarrow 0$ is equivalent to studying the behaviour of $n_{\e}(t,{\bf r},x)$ over many cell generations and in the case where spontaneous epimutations and random cell movement induce small changes in the local phenotypic distribution. 

Moreover, in agreement with much of the previous work on the mathematical analysis of the evolutionary dynamics of continuously-structured populations~\cite{perthame2006transport}, we consider the case where at time $t=0$ tumour cells that occupy the same position are mainly in the same phenotypic state, that is, at every position ${\bf r}$ the initial local cell phenotypic distribution $n_{\e}(0,{\bf r},x)$ is a sharp Gaussian-like function with mean value $\bar{x}^0({\bf r})$ and integral $I_{\e}(0,{\bf r})$. Hence, we assume
\begin{equation}
\label{ice0}
n_{\e}(0,{\bf r},x) = e^{\ue^0({\bf r},x)/\e}
\end{equation}
with $\ue^0({\bf r},x)$ being a smooth, uniformly concave function of $x$ for every ${\bf r} \in \Omega$ such that
\begin{equation}
\label{icIeb}
0 < I_{\e}(0,{\bf r}) < \infty 
\end{equation}
and
\begin{equation}
\label{ice0b}
e^{\ue^0({\bf r},x)/\e} \xrightharpoonup[\e  \rightarrow 0]{\scriptstyle\ast} I(0,{\bf r}) \, \delta_{\bar{x}^0({\bf r})}(x) \quad \text{for all } {\bf r} \in \Omega
\end{equation}
in the sense of measures, where $\delta_{\bar{x}^0({\bf r})}(x)$ is the Dirac delta distribution centred at $\bar{x}^0({\bf r})$.

Building upon the method presented in~\cite{barles2009concentration,diekmann2005dynamics,lorz2011dirac,perthame2006transport,perthame2008dirac}, we make the real phase WKB ansatz~\cite{barles1989wavefront,evans1989pde,fleming1986pde}
\begin{align}\label{eq3}
\nep(t,{\bf r},x) = e^{\ue(t,{\bf r},x)/\e}.
\end{align}
Substituting this ansatz into~\eqref{eq:ne} and using 
\begin{align*}  
&\dt\nep = \e^{-1}\nep\dt\ue, \quad \dr\nep = \e^{-1}\nep\dr\ue, \quad \dx\nep = \e^{-1}\nep\dx\ue, \\
&\Delta_{\bf r}\nep = \Big(\e^{-1} |\dr\ue|\Big)^{2}\nep + \e^{-1}\nep\Delta_{\bf r}\ue,\quad \dxx\nep = \Big(\e^{-1}\dx\ue\Big)^{2}\nep + \e^{-1}\nep\dxx\ue,\
\end{align*}
we obtain 
\begin{align}\label{eq4}
\dt\ue = R\big(x,I_{\e}(t,{\bf r}),s^{\infty}({\bf r})\big) + (\dx\ue)^2 + |\dr\ue|^2 +\e \, \left(\dxx\ue + \Delta_{\bf r}\ue \right)
\end{align}
subject to the initial condition $\ue(0,{\bf r},x)=\ue^0({\bf r},x)$, with $\ue^0({\bf r},x)$ given by~\eqref{ice0}.  

Letting $\e \to 0$ in~\eqref{eq4} we formally obtain the following equation for the leading-order term $u$ of the asymptotic expansion for $\ue$
\begin{align}\label{eq4eto0}
\dt u = R(x,I(t,{\bf r}),s^{\infty}({\bf r})) + (\dx u)^2 + |\dr u|^2 \quad \text{in } \Omega \times (0,1),
\end{align}
where $I(t,{\bf r})$ is the leading-order term of the asymptotic expansion for $I_{\e}(t,{\bf r})$. 

Since $\ue^0$ is a uniformly concave function of $x$ and, under assumption~\eqref{assp}, $R$ is a concave function of $x$ as well, we expect $u$ to be a concave function of $x$~\cite{barles2009concentration,mirrahimi2015asymptotic,perthame2008dirac}. Therefore, we formally have that there exists a unique locally dominant phenotypic state $\bar{x}(t,{\bf r})$, which is such that $\displaystyle{u(t,{\bf r},\bar{x}(t,{\bf r})) = \max_{x \in [0,1]} u(t,{\bf r},x)}$ and 
\beq
\label{dxu0}
\dx u(t,{\bf r},\bar{x}(t,{\bf r})) =0.
\eeq
Moreover, \eqref{assp0} and~\eqref{icIeb} ensure that $0 < I_{\e}(t,{\bf r}) < \infty$ and, therefore, letting $\e \to 0$ in~\eqref{eq3} gives the constraint 
\beq
\label{u0}
u(t,{\bf r},\bar{x}(t,{\bf r})) = 0 \quad \text{for all } (t,{\bf r}) \in (0,\infty) \times \Omega.
\eeq
Evaluating~\eqref{eq4eto0} at $x = \bar{x}(t,{\bf r})$ and using~\eqref{dxu0} along with~\eqref{u0} yields 
\beq
\label{Req0}
R(\bar{x}(t,{\bf r}),I(t,{\bf r}),s^{\infty}({\bf r})) = 0.
\eeq
Differentiating~\eqref{dxu0} with respect to $t$ yields
$$
\partial_{tx} u(t,{\bf r},\bar{x}(t,{\bf r})) +\dxx u (t,{\bf r},\bar{x}(t,{\bf r})) \, \dt \bar{x}(t,{\bf r})=0
$$
and, using the fact that $\dxx u (t,{\bf r},\bar{x}(t,{\bf r})) < 0$, we can formally rewrite the above equation as
\beq
\label{canoneqepspre}
\dt \bar{x}(t,{\bf r}) = -(\dxx u(t,{\bf r},\bar{x}(t,{\bf r}))^{-1} \partial_{tx} u(t,{\bf r},\bar{x}(t,{\bf r})).
\eeq
Furthermore, differentiating both sides of~\eqref{eq4eto0} with respect to $x$, evaluating the resulting equation at $x=\bar{x}(t,{\bf r})$ and using~\eqref{dxu0} along with~\eqref{u0} gives
$$
\partial_{tx}u(t,{\bf r},\bar{x}(t,{\bf r})) = \dx R\big(\bar{x}(t,{\bf r}),I(t,{\bf r}),s^{\infty}({\bf r})\big).
$$
Substituting the latter equation into~\eqref{canoneqepspre} we formally obtain the following equation for $\bar{x}(t,{\bf r})$
\beq
\label{canon}
\dt\xb(t,{\bf r}) = -(\dxx u(t,{\bf r},\xb(t,{\bf r})))^{-1} \, \dx R(\xb(t,{\bf r}),I(t,{\bf r}),s^{\infty}({\bf r})).
\eeq

Combining~\eqref{Req0} and~ \eqref{canon} we find that the steady-state values of $I(t,{\bf r})$ and $\xb(t,{\bf r})$, say $\ii({\bf r})$ and $\xb^{\infty}({\bf r})$, need to satisfy 
$$
\begin{cases}
R(\xbi({\bf r}),\ii({\bf r}),s^{\infty}({\bf r}))=0, 
\\\\
\dx R(\xbi({\bf r}),\ii({\bf r}),s^{\infty}({\bf r}))=0.
\end{cases}
$$
Substituting~\eqref{def:R} into the above system of equations,
we formally obtain
\beq
\begin{cases}
p(\xbi({\bf r}),s^{\infty}({\bf r})) - \kappa \ii({\bf r}) =0, 
\\\\
\dx p(\xbi({\bf r}),s^{\infty}({\bf r}))=0,  
\end{cases}
\;
\Longrightarrow
\;
\begin{cases}
\displaystyle{\ii({\bf r}) = \frac{1}{\kappa} \, p(\xbi({\bf r}),s^{\infty}({\bf r}))}, 
\\\\
\displaystyle{\xbi({\bf r}) = \argmax_{x \in [0,1]}  p(x,s^{\infty}({\bf r}))}.  
\end{cases}
\label{eq10}
\eeq
Taken together, these formal results indicate that, in the framework of the assumptions considered in this section, we can expect the local cell phenotypic distribution at steady-state $n^{\infty}({\bf r},x)$ to be of the form
\beq
\label{ninfty}
n^{\infty}({\bf r},x) = \ii({\bf r}) \, \delta_{\xbi({\bf r})}(x),
\eeq
with the local cell density $\ii({\bf r})$ 
and the locally dominant phenotypic state $\xbi({\bf r})$ both given by~\eqref{eq10}. This also implies that the local mean phenotypic state of the tumour cells at steady-state, say $X^{\infty}({\bf r})$, coincides with $\xbi({\bf r})$, that is,
$$
X^{\infty}({\bf r}) := \frac{1}{\ii({\bf r})} \int_{0}^1 x \, n^{\infty}({\bf r},x) \, {\rm d}x = \xbi({\bf r}).
$$

\section{Numerical solutions}
\label{sec:numsol}
In this section, we construct numerical solutions to the problem given by~\eqref{eq:n} and~\eqref{eq:s} subject to suitable initial conditions. First, we describe the setup of numerical simulations and the model parameterisation (see Section~\ref{setupnum}). Then, we present a sample of numerical solutions that confirm the results of the formal analysis of evolutionary dynamics carried out in Section~\ref{sec:analysis}. We consider both the case of an arbitrary distribution of blood vessels (see Section~\ref{illustratinganalysis}) and the case where the blood vessel distribution is reconstructed from clinical images obtained via D-OCT (Section~\ref{clinical}). Finally, we use numerical solutions of the model equations to assess the impact of tumour tissue vascularisation on intratumour phenotypic heterogeneity (see Section~\ref{numsimhet}).

\subsection{Setup of numerical simulations and model parameterisation}
\label{setupnum}
Building upon the modelling strategy presented in~\cite{lorenzi2018role}, we define the functions $f(x)$ and $g(x,s)$ that compose the cell division rate $p(x,s)$ as
\beq
\label{def:fg}
f(x) := \varphi \; \Big[1 - (1-x)^2\Big] \quad \text{and} \quad g(x,s) := \gamma \frac{s}{\mu + s} \; \big(1-x^2\big), \quad  \varphi < \gamma.
\eeq
In~\eqref{def:fg}, the parameters $\varphi>0$ and $\gamma>0$ model the maximum rate of cell division via anaerobic and aerobic pathways, respectively, while $\mu>0$ is the Michaelis-Menten constant of oxygen. The assumption $\varphi < \gamma$ is based on empirical evidence indicating that higher levels of expression of hypoxia-inducible factors correlate with lower rates of cell division~\cite{brown1998unique,brown2004exploiting,lloyd2016darwinian}. 
Definitions~\eqref{def:fg} satisfy the general assumptions~\eqref{assf}-\eqref{assg2} and \eqref{assp}, and allow for a detailed quantitative characterisation of the evolutionary dynamics of tumour cells based on the formal results presented in Section~\ref{sec:analysis}. Moreover, they lead to a phenotypic fitness landscape of the tumour $R(x,I,s)$ that is close to the approximate fitness landscapes which can be inferred from experimental data through regression techniques~\cite{lande1983measurement, otwinowski2014inferring,stinchcombe2008estimating}. In fact, substituting~\eqref{def:fg} into~\eqref{def:p}, after a little algebra we find
\beq
\label{pcomp} 
p(x,s)= a(s) - b(s) \left(x - h(s) \right)^2,
\eeq
where
\beq
\label{ab} 
a(s) := \gamma \dfrac{s}{\mu + s} + \frac{\varphi^2}{\varphi+\gamma \dfrac{s}{\mu + s}}, \quad b(s) := \varphi + \gamma \dfrac{s}{\mu + s}, \quad h(s) := \frac{\varphi}{\varphi+\gamma \dfrac{s}{\mu + s}}.
\eeq
Here, $a(s(t,{\bf r}))$ is the maximum fitness, $h(s(t,{\bf r}))$ is the fittest phenotypic state and $b(s(t,{\bf r}))$ is a nonlinear selection gradient at position ${\bf r}$ and time $t$ under the environmental conditions corresponding to the oxygen concentration $s(t,{\bf r})$.

We let the rate of inflow of oxygen through intratumour blood vessels to be constant in time and the same for all vessels, \emph{i.e.} we define the function $S(t,{\bf r})$ in~\eqref{q} as
\beq
\label{constS}
S(t,{\bf r}) \equiv S_v >0.
\eeq
Moreover, we consider different definitions of the set $\omega$ which represent different distributions of the blood vessels, as detailed in the next subsections. Unless otherwise explicitly stated, numerical simulations are carried out using the parameter values listed in Table~\ref{Tab1}, which are chosen to be consistent with the existing literature.

We define $\Omega := [0,{\rm L}] \times [0,{\rm L}]$ and choose ${\rm L}=0.5$. Under the parameter choice of Table~\ref{Tab1}, this value of ${\rm L}$ is equivalent to considering a square region of a 2D cross-section of a vascularised tumour of area $2.5 \times 10^{-3} \, cm^2$. Moreover, we assume $t \in [0,{\rm T}]$, with the final time ${\rm T}$ such that the numerical solutions are sufficiently close to equilibrium at the end of simulations. Finally, we use the notation ${\bf r} \equiv (r_1,r_2)$. 

We complement~\eqref{eq:n} and~\eqref{eq:s} with the following initial conditions
\beq
\label{ic}
n(0,{\bf r},x) = 10^{8}\,e^{-\frac{(x-0.5)^2}{0.1}} \; \forall \, {\bf r} \in \Omega \quad \text{and} \quad s(0,{\bf r}) = S_0 \, {\bf 1}_{\omega}({\bf r}),
\eeq
where $S_0 = 6.3996 \times 10^{-7}$ $g \, cm^{-2}$~\cite{kumosa2014permeability}. The initial conditions~\eqref{ic} correspond to a biological scenario whereby at time $t=0$ tumour cells are uniformly distributed across $\Omega$ and are mainly in the intermediate phenotypic state $x=0.5$, while the oxygen is concentrated in correspondence of the blood vessels. 

Numerical solutions are constructed using a uniform discretisation of the square $[0,{\rm L}] \times [0,{\rm L}]$ as the computational domain of the independent variable ${\bf r}$ and a uniform discretisation of the set $[0,1]$ as the computational domain of the independent variable $x$. We also discretise the interval $[0,{\rm T}]$ with a uniform step. The method for constructing numerical solutions is based on an explicit finite difference scheme in which a three-point and a five-point stencils are used to approximate the diffusion terms in $x$ and ${\bf r}$, respectively, and an explicit finite difference scheme is used for the reaction terms~\cite{leveque2007finite}. All numerical computations are performed in {\sc Matlab}. 
\begin{table*}[tbhp]
	{\footnotesize
		\centering
		\caption{Parameter values used in numerical simulations}
		\hspace{+1cm}\begin{tabular}{cllc}
			{\it Parameter} & {\it Biological meaning} & {\it Value} & {\it Reference} \\
			$\mu$ & Michaelis-Menten constant of oxygen & $1.5 \times 10^{-7}$ $g \, cm^{-2}$ & \cite{casciari1992variations} \\ 
			$\beta$ & Cell motility & $10^{-13}$ $cm^2 \, s^{-1}$ & \cite{smith2004measurement,wang2009mathematical} \\ 
			$\beta_s$ & Oxygen diffusivity & $2 \times 10^{-5}$ $cm^2 \, s^{-1}$ & \cite{hlatky1985two} \\
			$\gamma$ & Maximum rate of cell division via aerobic pathways & $1 \times 10^{-4}$ $s^{-1}$ & \cite{casciari1992variations, ward1997mathematical} \\ 
			$\kappa$ & Rate of cell death due to competition for space  & $2 \times 10^{-13}$ $cm^{2} \, s^{-1} \, cells^{-1}$ & \cite{li1982glucose,lorenzi2018role}  \\ 
			$\eta_s$ & Conversion factor for cell consumption of oxygen & $2 \times 10^{-11}$ $g \, cells^{-1}$ & \cite{casciari1992variations,lorenzi2018role} \\
			$\alpha$ & Rate of spontaneous phenotypic changes & $10^{-13}$ $s^{-1}$ & \cite{doerfler2006dna,duesberg2000explaining} \\ 
			$\lambda_s$ & Rate of natural decay of oxygen & $2.78 \times 10^{-6}$ $s^{-1}$& \cite{cumsille2015proposal}  \\ 
			$\varphi$ & Maximum rate of cell division via anaerobic pathways & $1 \times 10^{-5}$ $s^{-1}$ & \cite{gordan2007hif} \\ 
			$S_v$ & Constant rate of inflow of oxygen through blood vessels & $6.3996 \times 10^{-7}$ $g \, cm^{-2} \, s^{-1}$ & \cite{kumosa2014permeability}   
		\end{tabular}
		\label{Tab1}
	}
\end{table*}

\subsection{Numerical solutions for arbitrary blood vessel distributions}
\label{illustratinganalysis}
Substituting~\eqref{pcomp} into~\eqref{eq10} gives the following expressions for the local cell density and the locally dominant phenotypic state at steady-state
\begin{equation}\label{eq12}
I^{\infty}({\bf r}) \equiv I^{\infty}(s^{\infty}({\bf r}))  = \frac{a(s^{\infty}({\bf r}))}{\kappa} \quad \text{and} \quad \bar{x}^{\infty}({\bf r}) \equiv \bar{x}^{\infty}(s^{\infty}({\bf r})) = h(s^{\infty}({\bf r})),
\end{equation}
where $a(s^{\infty})$ and $h(s^{\infty})$ are the maximum fitness and the fittest phenotypic state defined according to~\eqref{ab}.

\begin{figure}[htb!]
\centering
\includegraphics[width=0.9\linewidth]{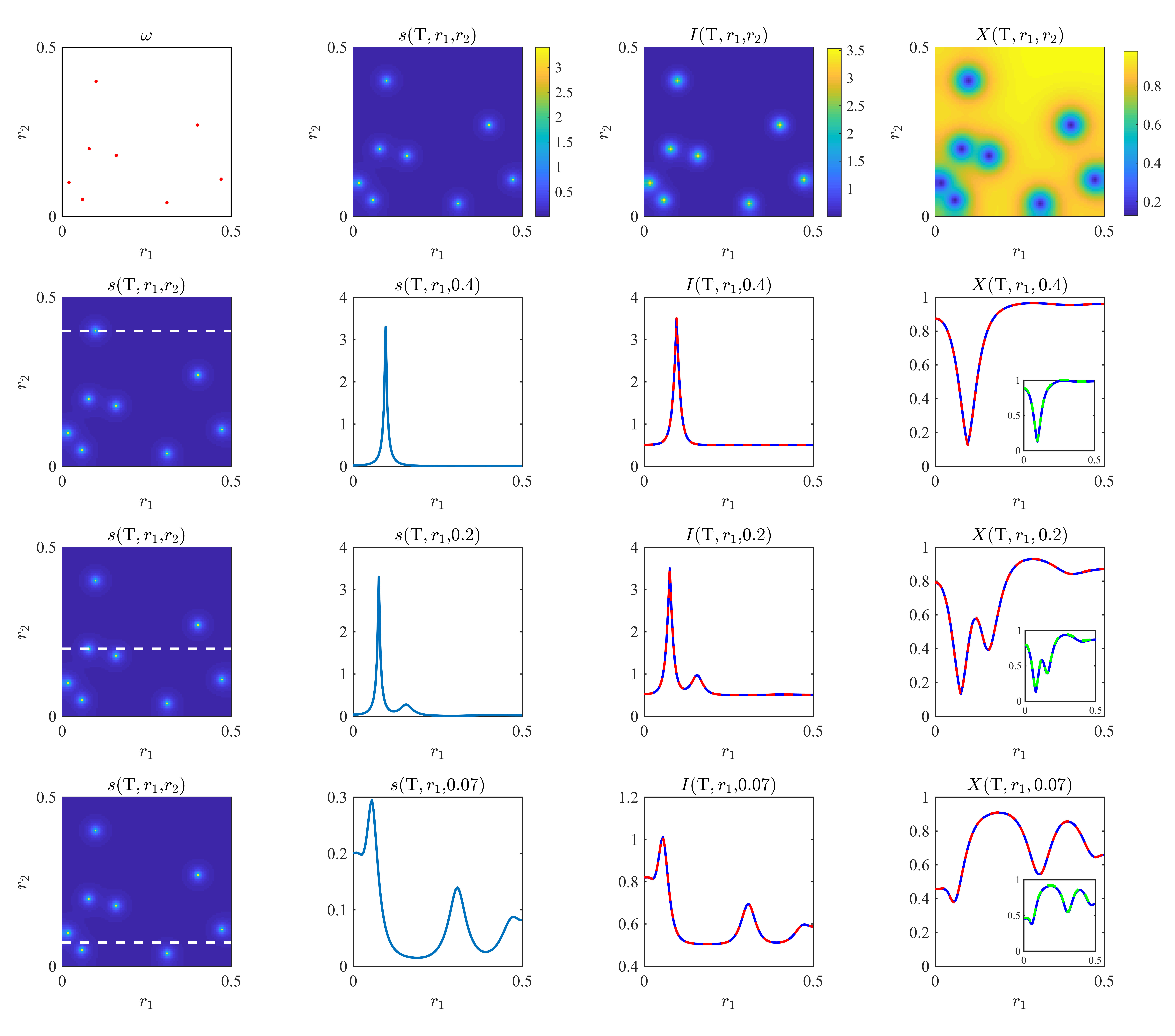} 
\caption{{\bf First row.} Plots of the oxygen concentration $s({\rm T}, {\bf r})$ (second panel), the cell density $I({\rm T}, {\bf r})$ (third panel) and the mean phenotypic state $X({\rm T}, {\bf r})$ (fourth panel) obtained by solving numerically the problem given by~\eqref{eq:n} and~\eqref{eq:s} subject to the initial conditions~\eqref{ic}, with the set $\omega$ corresponding to the parts of $\Omega$ highlighted in red in the first panel and with the parameter values listed in Table~\ref{Tab1}. {\bf Second row.} Plots of the oxygen concentration $s({\rm T},r_1, 0.4)$ (second panel), the cell density $I({\rm T}, r_1, 0.4)$ (third panel, blue line) and the mean phenotypic state $X({\rm T}, r_1, 0.4)$ (fourth panel, blue line). The plot of the oxygen distribution $s({\rm T}, {\bf r})$ is displayed in the first panel, where the white, dashed line highlights the 1D cross-section corresponding to $r_2=0.4$. The red lines in the third and fourth panels highlight $I^{\infty}({\bf r})$ and $\bar{x}^{\infty}({\bf r})$ computed through~\eqref{eq12} choosing $s^{\infty}({\bf r})=s({\rm T},{\bf r},x)$. The inset in the fourth panel displays the plot of the mean phenotypic state $X({\rm T}, r_1, 0.4)$ (blue line) and of the maximum point of $n({\rm T}, r_1, 0.4,x)$ (green line). {\bf Third and fourth row.} Same as the second row but for $r_2=0.2$ (third row) and $r_2=0.07$ (fourth row). The oxygen concentration $s({\rm T}, {\bf r})$ is in units of $10^{-7} \, g \, cm^{-2}$, the cell density $I({\rm T}, {\bf r})$ is in units of $10^8 \, cells \, cm^{-2}$, and the spatial variables $r_1$ and $r_2$ are in units of $cm$.} \label{random}
\end{figure}

In agreement with the results of the formal analysis of evolutionary dynamics carried out in Section~\ref{sec:analysis} [\emph{cf.} the expression~ \eqref{ninfty} of the local cell phenotypic distribution at steady-state], the numerical results displayed in Fig.~\ref{random} show that the cell density $I({\rm T},{\bf r})$ and the mean phenotypic state $X({\rm T},{\bf r})$, which are computed using the numerical solution $n({\rm T},{\bf r},x)$ of~\eqref{eq:n}, coincide with $I^{\infty}({\bf r})$ and $\bar{x}^{\infty}({\bf r})$, which are computed through~\eqref{eq12} choosing $s^{\infty}({\bf r})=s({\rm T},{\bf r},x)$, where $s({\rm T},{\bf r},x)$ is the numerical solution of~\eqref{eq:s}. Moreover, the numerical solution $n({\rm T},{\bf r},x)$ is concentrated as a sharp Gaussian-like function (data not shown) with maximum at the mean phenotypic state $X({\rm T},{\bf r})$ (\emph{vid.} insets in the fourth panels of Fig.~\ref{random}).

The numerical results of Fig.~\ref{random}, which refer to the blood vessel distribution displayed in the first panel of the figure, indicate that the oxygen concentration is maximal in the vicinity of the blood vessels and decreases monotonically with the distance from the blood vessels. Accordingly, the cell density is higher in the regions in close proximity to the blood vessels and the mean phenotypic state increases from values close to $x=0$ (\emph{i.e.} low levels of expression of the hypoxia-inducible factor) to values close to $x=1$ (\emph{i.e.} high levels of expression of the hypoxia-inducible factor) moving away from the blood vessels. These results communicate the biological notion that spatial inhomogeneities in the distribution of oxygen -- which emerge spontaneously as a result of the nonlinear interplay between the spatial distribution of the blood vessels, the reaction-diffusion dynamics of the oxygen and the consumption of oxygen by the cells -- create environmental conditions that favour the selection of cells with different phenotypic characteristics in different regions of vascularised tumours, thus leading to the emergence and development of intratumour phenotypic heterogeneity.   


\subsection{Numerical solutions for blood vessel distributions reconstructed from clinical images}
\label{clinical}
\begin{figure}[h!]
    \centering
    \includegraphics[width=0.95\linewidth]{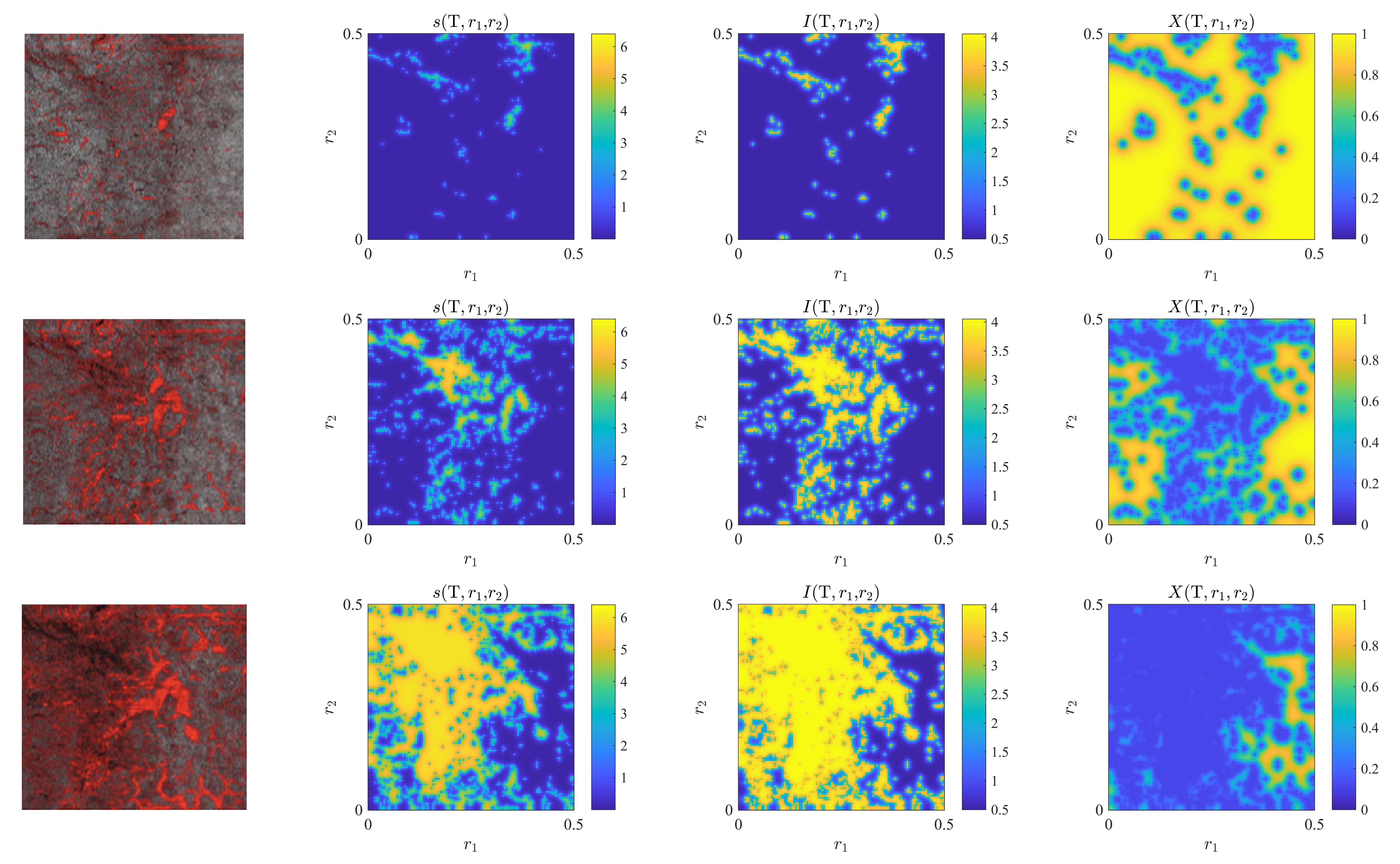}
    \caption{{\bf First row.} Plots of the oxygen concentration $s({\rm T}, {\bf r})$ (second panel), the cell density $I({\rm T}, {\bf r})$ (third panel) and the mean phenotypic state $X({\rm T}, {\bf r})$ (fourth panel) obtained by solving numerically the problem given by~\eqref{eq:n} and~\eqref{eq:s} subject to the initial conditions~\eqref{ic}, with the set $\omega$ reconstructed from the blood vessel distribution provided by the clinical image displayed in the first panel, where the intratumoural vascular network is highlighted in red, and with the parameter values listed in Table~\ref{Tab1} except for $\eta_s=2\times10^{-10}$  $g \, cell^{-1}$. 
{\bf Second and third row.} Same as the first row but for a different clinical image. Clinical images are taken from~\cite[Fig. 5(d-f)]{schuh2017imaging} under Creative Commons licence \url{https://creativecommons.org/licenses/by-nc/4.0/}. These images correspond to three cross sections of a malignant melanoma at a depth of $0.02 \, cm$ (first row), $0.03 \, cm$ (second row) and $0.04 \, cm$ (third row) from the surface of the epidermis. The oxygen concentration $s({\rm T}, {\bf r})$ is in units of $10^{-7} \, g \, cm^{-2}$, the cell density $I({\rm T}, {\bf r})$ is in units of $10^8 \, cells \, cm^{-2}$, and the spatial variables $r_1$ and $r_2$ are in units of $cm$.}
    \label{experimental}
\end{figure}

The plots in Fig.~\ref{experimental} demonstrate that the qualitative behaviour of the numerical results in Fig.~\ref{random} remains unchanged when spatial distributions of the intratumour blood vessels reconstructed from clinical images are considered. These are the plots of the oxygen concentration $s({\rm T}, {\bf r})$, the cell density $I({\rm T}, {\bf r})$ and the mean phenotypic state $X({\rm T}, {\bf r})$ obtained by solving numerically the problem given by~\eqref{eq:n} and~\eqref{eq:s} subject to the initial conditions~\eqref{ic}, with $\omega$ defined according to the distributions of blood vessels provided by the clinical images displayed in the first column of the figure, which were obtained via D-OCT and correspond to three cross sections of a malignant melanoma at a depth of $0.02 \, cm$ (top panel), $0.03 \, cm$ (central panel) and $0.04 \, cm$ (bottom panel) from the surface of the epidermis~\cite[Fig. 5]{schuh2017imaging}. 
These results also indicate that increasing levels of tumour vascularisation (from top to bottom panel in the first column) lead to a more homogeneous spatial distribution of oxygen (second column), which correlates with a more uniform cell density (third column) and a less diverse mean phenotypic state (fourth column). This suggests the existence of a relationship between the level of tumour tissue vascularisation and the level of intratumour phenotypic heterogeneity, which is systematically investigated in the next subsection.  


\subsection{Numerical solutions to assess the impact of tumour tissue vascularisation on intratumour phenotypic heterogeneity}
\label{numsimhet}
In order to systematically assess the impact of tumour tissue vascularisation on the level of intratumour phenotypic heterogeneity, we carry out numerical simulations considering first increasing numbers of regularly distributed blood vessels, which correspond to increasing values of the vascular density $\varrho$ defined as
\begin{equation}
\label{density}
\displaystyle{\varrho := \frac{|\omega|}{|\Omega|}},
\end{equation}
{and then different random distributions of blood vessels characterised by increasing levels of vessel clustering for a fixed vascular density. We quantify the level of intratumour phenotypic heterogeneity through the following continuum versions of the equitability index $E(t)$ (defined as a rescaled Shannon diversity index) and the Simpson diversity index $D(t)$~\cite{shannon1948mathematical,simpson1949measurement} 
\begin{equation}\label{equitabilitysimpson}
E(t) := - \int_0^1 \frac{F(t,x) \, \log{F(t,x)}}{\log{N(t)}} \, {\rm d}x \quad \text{and} \quad D(t) := \left(\int_0^1 F^2(t,x) \, {\rm d}x \right)^{-1},
\end{equation}
where the total cell number $N(t)$ and the fraction $F(t,x)$ of cells in the phenotypic state $x$ within the tumour are defined according to~\eqref{NF}. 

The results obtained varying the vascular density $\varrho$ are summarised by the plots in Fig.~\ref{heterogeneity}, which display the equitability index and the Simpson diversity index at the end of numerical simulations as functions of $\varrho$. Both diversity indices are relatively low for small values of the vascular density, increase and reach a maximum value for intermediate values of the vascular density -- notice that both $E({\rm T})$ and $D({\rm T})$ attain their maximum at the same value of $\varrho$ -- and then decrease again for high values of the vascular density. This is due to the fact that, as shown by the insets in Fig.~\ref{heterogeneity}: for low blood vessel densities the oxygen concentration $s({\rm T}, {\bf r})$ is uniformly low throughout $\Omega$ and, therefore, the mean phenotypic state $X({\rm T}, {\bf r})$ is uniformly close to $x=1$ ({\it cf.} the insets related to $\varrho=0.4\times 10^{-3}$); for intermediate blood vessel densities the oxygen concentration is more heterogeneously distributed and, as a consequence, the mean phenotypic state is more diverse ({\it cf.} the insets related to $\varrho=2.5\times 10^{-3}$); for high blood vessel densities the oxygen concentration is relatively high throughout the tumour tissue and the mean phenotypic state is on average close to $x=0$ ({\it cf.} the insets related to $\varrho=8.1\times 10^{-3}$).
\begin{figure}[h!] 
    \centering
	\includegraphics[width=0.9\linewidth]{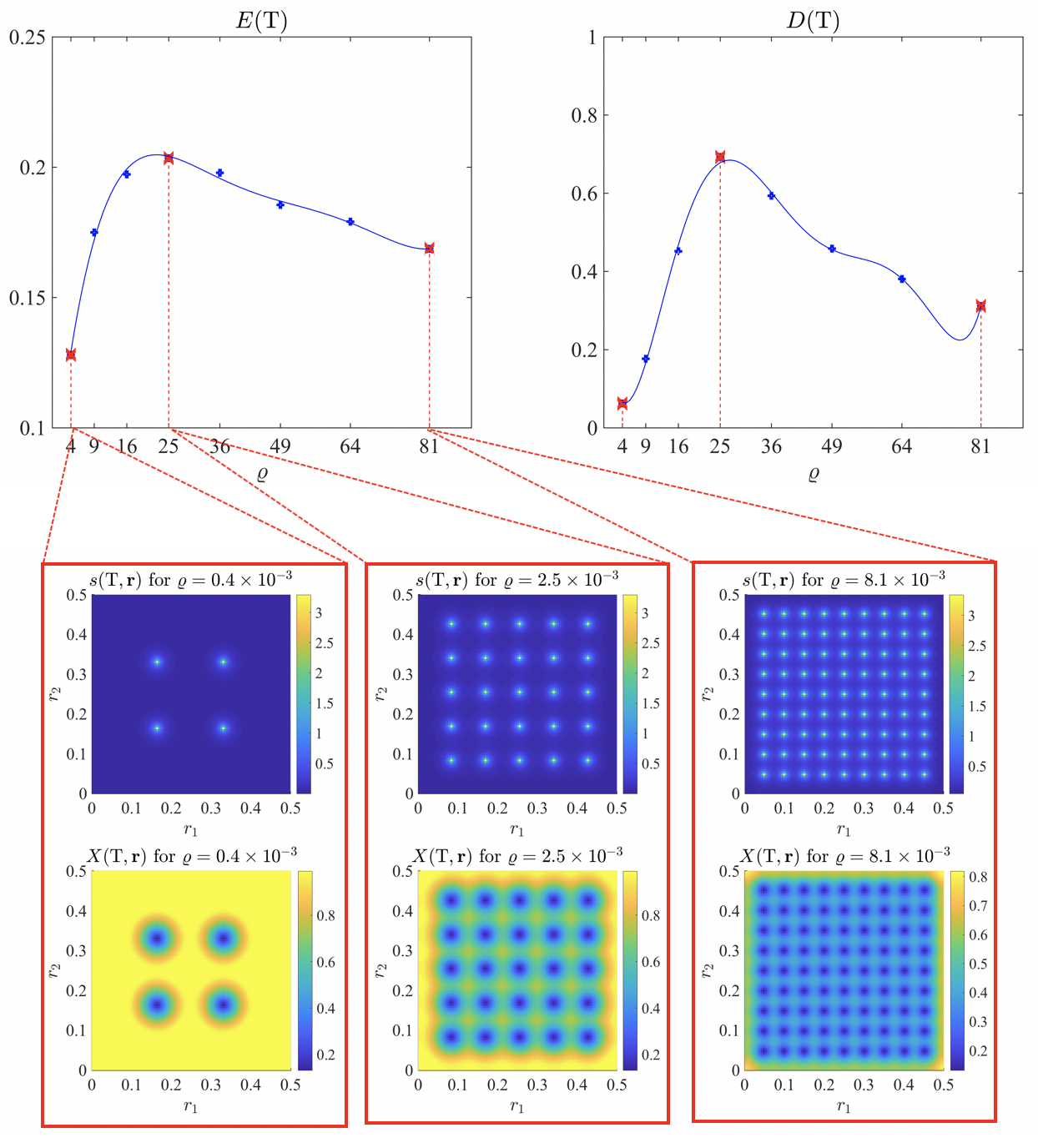}
	\caption{\label{heterogeneity} Plots of the equitability index $E({\rm T})$ and the Simpson diversity index $D({\rm T})$ for different definitions of the set $\omega$ characterised by different values of the vascular density $\varrho$ defined according to~\eqref{density}. The equitability index and the Simpson diversity index are computed numerically through formulas~\eqref{equitabilitysimpson} using the numerical solutions of the problem given by~\eqref{eq:n} and~\eqref{eq:s} subject to the initial conditions~\eqref{ic}, with the parameter values listed in Table~\ref{Tab1}. The insets display sample plots of the oxygen distributions $s({\rm T}, {\bf r})$ (top panel) and the mean phenotypic state $X({\rm T}, {\bf r})$ (bottom panel) corresponding to different values of $\varrho$. The Simpson diversity index $D({\rm T})$ is in units of $10^4$, the vascular density $\varrho$ is in units of $10^{-4}$, the oxygen concentration $s({\rm T}, {\bf r})$ is in units of $10^{-7} \, g \, cm^{-2}$, and the spatial variables $r_1$ and $r_2$ are in units of $cm$.} 
\end{figure}

\clearpage

The results obtained varying the level of blood vessel clustering for a fixed vascular density $\varrho$ are summarised by the plots in Fig.~\ref{heterogeneity_random}, which display the oxygen distribution $s({\rm T}, {\bf r})$ and the mean phenotypic state $X({\rm T}, {\bf r})$, along with the corresponding fraction of cells in each phenotypic state $F({\rm T},x)$ and diversity indices $E({\rm T})$ and $D({\rm T})$. These results refer to an intermediate value of $\varrho$ that corresponds to the maximum of the equitability index and the Simpson diversity index displayed in Fig.~\ref{heterogeneity} (\textit{i.e.} $\varrho=25\times 10^{-4}$). Both diversity indices decrease as the level of blood vessel clustering increases ({\it cf.} the values of $E({\rm T})$ and $D({\rm T})$ in the insets of the panels in the third column of Fig.~\ref{heterogeneity_random}). In fact, for lower levels of blood vessel clustering the oxygen concentration $s({\rm T}, {\bf r})$ is more heterogeneously distributed and, as a consequence, the mean phenotypic state $X({\rm T}, {\bf r})$ is more diverse and the cell phenotypic distribution across $\Omega$ given by $F({\rm T},x)$ is rather uniform ({\it cf.} the plots in the first row of Fig.~\ref{heterogeneity_random}). On the other hand, for higher levels of blood vessel clustering, the oxygen concentration is relatively high in the regions in close proximity to the clusters of blood vessels and relatively low throughout the rest of tumour tissue. As a result, the mean phenotypic state is mostly close to $x=1$ with the exception of the regions near the clusters of blood vessels where it is close to $x=0$, and the cell phenotypic distribution across the whole tumour is approximatively bimodal, with a high peak at $x=1$ and a low peak at $x=0$ ({\it cf.} the plots in the third row of Fig.~\ref{heterogeneity_random}).

\begin{figure}[h!] 
    \centering
	\includegraphics[width=0.8\linewidth]{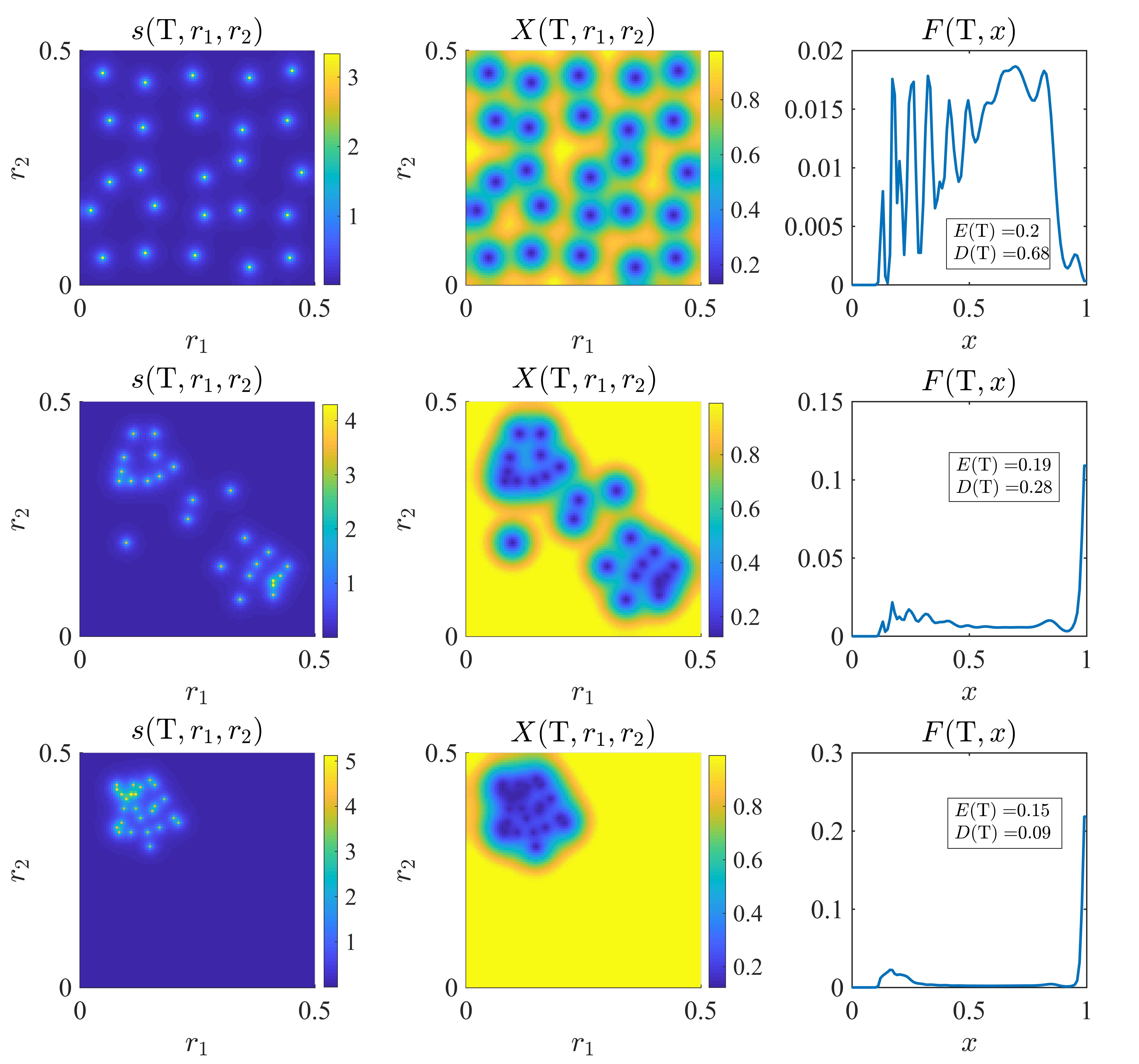}
	\caption{\label{heterogeneity_random} {\bf First row.} Plots of the oxygen distribution $s({\rm T}, {\bf r})$  (first panel), mean phenotypic state $X({\rm T}, {\bf r})$ (second panel) and fraction of cells in each phenotypic state $F({\rm T},x)$ defined via~\eqref{NF} (third panel), for a definition of the set $\omega$ corresponding to a random distribution of blood vessels characterised by vascular density $\varrho=25\times 10^{-4}$ and a low level of blood vessel clustering. The values of the corresponding equitability index $E({\rm T})$ and Simpson diversity index $D({\rm T})$, which are computed numerically through formulas~\eqref{equitabilitysimpson}, are provided in the inset of the third panel. {\bf Second and third row.} Same as the first row but for a definition of the set $\omega$ corresponding to an intermediate level (second row) and a high level (third row) of blood vessel clustering. The oxygen concentration $s({\rm T}, {\bf r})$ is in units of $10^{-7} \, g \, cm^{-2}$, the spatial variables $r_1$ and $r_2$ are in units of $cm$, and the Simpson diversity index $D({\rm T})$ is in units of $10^4$. }
\end{figure}

\section{Conclusions and research perspectives}
\label{sec:disc}
Intratumour phenotypic heterogeneity poses a major obstacle to anti-cancer therapy~\cite{burrell2014tumour,chisholm2015emergence,gillies2012evolutionary,lipinski2016cancer,michor2010origins,shah2001cell}. It has been hypothesised that the emergence of phenotypic heterogeneity among cancer cells within malignant tumours is an eco-evolutionary process driven by spatial variability in the distribution of abiotic factors, which supports the creation of distinct ecological niches whereby cells with different phenotypic characteristics can be selected~\cite{alfarouk2013riparian,kaznatcheev2017cancer,marusyk2012intra,sun2015intra}. In particular, oxygen is one of the key abiotic components of the tumour microenvironment that are implicated in the emergence of intratumour phenotypic heterogeneity~\cite{gillies2012evolutionary,lorenzi2018role,sun2015intra}.

In this paper, we have undertaken a mathematical study of the eco-evolutionary dynamics of tumour cells within vascularised tumours. Our study is based on formal asymptotic analysis and numerical simulations of a non-local PDE model that describes the phenotypic evolution of tumour cells and their nonlinear dynamic interactions with the oxygen, which is released from the intratumoural vascular network. 

Our formal analytical results recapitulate the outcomes of previous theoretical and experimental studies on intratumour phenotypic heterogeneity~\cite{anderson2006tumor,gallaher2013evolution,gillies2012evolutionary,ibrahim2017defining} by providing a mathematical formalisation of the idea that local variations in the oxygen concentration bring about local variations in the phenotypic fitness landscape of the tumour, which ultimately result in a heterogeneous intratumour phenotypic composition. Numerical simulations of the model corroborate this conclusion and show that local variations in the oxygen concentration, which are orchestrated by nonlinear dynamic interactions between tumour cells and oxygen, promote the selection of cells in different phenotypic states within the tumour depending on the distance from the blood vessels. In particular, our results offer a theoretical basis for the biological evidence that regions of the tumour tissue in the vicinity of blood vessels are densely populated by proliferative phenotypic variants, while poorly oxygenated regions are sparsely populated by hypoxic phenotypic variants~\cite{padhani2007imaging,semenza2003targeting,sun2015intra,tannock1968relation}.     

Furthermore, the results of numerical simulations of the model establish a relation between the degree of tissue vascularisation and the level of intratumour phenotypic heterogeneity, measured either as the equitability index or the Simpson diversity index. This supports the idea that maps of the intratumour vascular network, which can be reconstructed from clinical images obtained via non-invasive imagine techniques, such as D-OCT~\cite{lavina2016brain,schuh2017imaging} and many others \cite{anderson2001measuring,fukumura2010tumor,grimes2016estimating,nobre2018different,padhani2007imaging}, could be  clinically relevant, as they could be used to inform targeted anticancer therapy~\cite{marusyk2012intra,powathil2012modeling,powathil2012modelling,vaupel2004tumor}.

Whilst we carried out numerical simulations considering a region of tumour tissue of area $2.5 \times 10^{-3} \, cm^2$, which was chosen in agreement with clinical images provided in~\cite{schuh2017imaging}, and using parameter values that are derived from specific cancer datasets, given the robustness and structural stability of the results of formal asymptotic analysis presented here, we expect the conclusions of this study about the emergence of substantial intratumour phenotypic heterogeneity driven by eco-evolutionary processes at the cellular scale to hold when larger tumour regions and different cancer datasets are considered.

We conclude with a brief overview of possible research directions. In oder to disentangle and quantify the impact of different evolutionary parameters on the emergence and development of intratumour phenotypic heterogeneity, it would be useful to have exact solutions of~\eqref{eq:n} in the case where the intratumour phenotypic fitness landscape is defined according to~\eqref{def:R} and~\eqref{pcomp}. This could be done by generalising the method developed in~\cite{almeida2019evolution,chisholm2016evolutionary,lorenzi2016tracking} to construct exact solutions of non-local parabolic PDEs modelling the dynamics of well-mixed phenotype-structured populations to the case where spatial structure is included. 

In addition, along the lines of~\cite{scott2016spatial}, further investigations on a possible link between the topology of tumour vasculature and the level of intratumour phenotypic heterogeneity could be undertaken. Moreover, building upon the ideas presented in~\cite{ardavseva2019evolutionary,ardavseva2019dissecting}, it would be interesting to study the effect  on the evolutionary dynamics of tumour cells of fluctuations in the rate of oxygen inflow, which are known to influence intratumour phenotypic heterogeneity~\cite{gillies2018eco,marusyk2012intra,robertson2015impact}. It would also be interesting to include the effect of temporal changes in the spatial distribution of intratumoural blood vessels, which would make it possible to explore the influence of angiogenesis on the eco-evolutionary dynamics of tumour cells in vascularised tumours. In this regard, it is a known fact that cancer cells in hypoxic conditions produce and secrete proangiogenic factors which induce the formation of new blood vessels departing from existing ones. 

While the focus of this work has been on the impact of spatial variability in the oxygen concentration on the emergence of intratumour phenotypic heterogeneity, building on~\cite{fiandaca2020mathematical}, it would be interesting to extend the modelling framework used here to incorporate the effect of nonlinear dynamic interactions between tumour cells and other abiotic factors, such as glucose and lactate, that are known to influence the levels of intratumour phenotypic heterogeneity~\cite{gatenby2007cellular,gatenby2007glycolysis,gillies2007hypoxia,kaznatcheev2017cancer,manem2015modeling,molavian2009fingerprint,robertson2015impact,zhao2013targeting}.

Furthermore, although well suited to modelling the dynamics of large cell populations, PDE models like that considered here cannot capture adaptive phenomena that are driven by stochasticity in the evolutionary paths of single cells. Therefore, it would also be interesting to complement the results of our study with numerical simulations of corresponding individual-based models which track the evolutionary trajectories of single cells across a space of discrete phenotypic states, as similarly done in~\cite{ardavseva2020comparative,chisholm2016evolutionary,chisholm2015emergence,stace2020discrete}. This would make it possible to have a more precise description of the phenotypic evolution of tumour cells in cases where cell numbers are relatively low and, therefore, stochastic fluctuations in single-cell phenotypic properties will have a stronger impact on intratumour phenotypic heterogeneity.

Finally, we plan to extend the model considered here to carry out a mathematical study of the eco-evolutionary dynamics of tumour cells in metastatic tumours. In this respect, a modelling approach analogous to the one presented in~\cite{franssen2019mathematical}, whereby different metastatic sites are represented as distinct compartments and the metastatisation process is modelled by allowing tumour cells to transition from one site to another through the intratumour blood vessels seen as entry/exit locations, may prove useful. 

\bigskip

\bibliographystyle{siam}
\bibliography{Bibliography}

\end{document}